\documentclass[prl,aps,superscriptaddress,showpacs,twocolumn]{revtex4-1}
\pdfoutput=1
\usepackage{graphicx}
\usepackage{amsmath}
\usepackage{graphicx}
\usepackage{color}
\usepackage{graphicx}
\usepackage{amssymb}
\usepackage{bm}
\newcommand{\hide}[1]{}

\newcommand{\eq}[1]{Eq.\,(\ref{#1})}

\newcommand{\fig}[1]{Fig.\,\ref{#1}}

\newcommand{\ket}[1]{\ensuremath{\left| #1 \right\rangle}}
\newcommand{\bra}[1]{\ensuremath{\left\langle #1 \right|}}

\newcommand{\mbf}[1]{\ensuremath{\mathbf{#1}}}

\begin{document}

\title{Ion crystal transducer for strong coupling between single ions and single photons}

\date{\today}

\author{L. Lamata}
\email{lucas.lamata@mpq.mpg.de. Current address: Universidad del Pa\'{\i}s Vasco, Bilbao, Spain.} \affiliation{Max-Planck-Institut
f\"ur Quantenoptik, Hans-Kopfermann-Strasse 1, 85748 Garching,
Germany}

\author{D. R. Leibrandt}
\email{Current address: NIST, Boulder, USA.}
\affiliation{Center for Ultracold Atoms, Department of Physics, MIT,
Cambridge, MA 02139, USA}

\author{I. L. Chuang}
\affiliation{Center for Ultracold Atoms, Department of Physics, MIT,
Cambridge, MA 02139, USA}

\author{J. I. Cirac}
\affiliation{Max-Planck-Institut f\"ur Quantenoptik,
Hans-Kopfermann-Strasse 1, 85748 Garching, Germany}

\author{M. D. Lukin}
\affiliation{ITAMP, Harvard-Smithsonian Center for Astrophysics,
Cambridge, MA 02138, USA} \affiliation{ Department of Physics,
Harvard University, Cambridge, MA 02138, USA}

\author{V. Vuleti\'{c}}
\affiliation{Center for Ultracold Atoms, Department of Physics, MIT,
Cambridge, MA 02139, USA}

\author{S. F. Yelin}
 \affiliation{ITAMP, Harvard-Smithsonian
Center for Astrophysics, Cambridge, MA 02138, USA}
\affiliation{Department of Physics, University of Connecticut,
Storrs, CT 06269, USA}

\begin{abstract}
A new approach for realization of a quantum interface between single
photons and single ions in an ion crystal is proposed and analyzed.
In our approach the coupling between a single photon and a single ion is
enhanced via the collective degrees of freedom of the ion crystal.
Applications including single-photon generation, a
memory for a quantum repeater, and a deterministic photon-photon,
photon-phonon, or photon-ion entangler are discussed.
\end{abstract}


\maketitle

 Realization of efficient quantum interfaces between single photons and single matter qubits is one of the
  most important and challenging goals in quantum information science~\cite{NielsenChuang}. It enables a wide variety of potential applications
  ranging from scalable quantum computing schemes to quantum networks and single-photon nonlinear optics~\cite{KLM}. Much
  progress has been achieved towards the realization of such quantum information implementations over the past decade
  \cite{Darkstatepolariton,Darkstatepol2,Darkstatepol3,Raman,Photon-echo,KimbleKuzmich,Kuzmich,Vuletic2,Kraus,Vuletic1,Gorshkov1,Gorshkov2,Gorshkov3,Drewsen,Review1,Review2}.
  Most of these have been based on neutral atoms, where quantum states  can be stored for long times
   \cite{Darkstatepolariton,Darkstatepol2,Darkstatepol3,Raman,Photon-echo,KimbleKuzmich,Kuzmich,Vuletic2,Kraus,Vuletic1,Gorshkov1,Gorshkov2,Gorshkov3}.
  The realization of a quantum optical interface for isolated single ions, which are among the most promising qubit
  candidates \cite{LeibfriedEtAl}, is still an outstanding challenge as the achievable coupling strength is typically small
  under realistic experimental conditions \cite{Blattion,Walther}.

\begin{figure}[h]
\includegraphics[width=.95\linewidth]{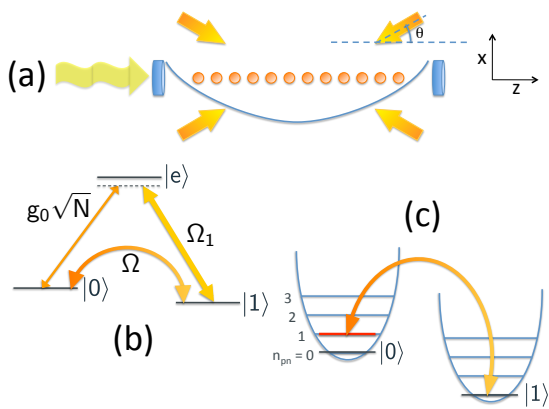}
\caption{Schematic of the setup: (a) Chain of $N$ ions in cavity with incoming photon and two pairs of off-axis lasers comprising a quadrupole field.
(b) Ionic level scheme: states $\ket{0}$ and
$\ket{1}$ are metastable states (e.g., $3D_{5/2}$ and $4S_{1/2}$ in
$^{40}$Ca$^+$), and $\ket{e}$ is an excited state, (e.g., the $4P_{3/2}$ state in the same ion).
 For step I, the transition
$\ket{0}\rightarrow\ket{e}$ is coupled by a single photon with the
effective Rabi frequency $g_0\sqrt{N}$. The $\Omega_1$ laser is directed perpendicularly to the chain. For step II, the states $\ket{0}$ and $\ket{1}$ are coupled by a quadrupole transition. (c) Coupling to a phonon
mode: the harmonic energy level of the phonon excitations, e.g.,
of the COM mode, is denoted for different internal states. The
singly excited phonon mode can be addressed by a transition
frequency that is red-detuned from the bare ionic transitions by
the frequency of the phonons, $\omega_{pn}$, by a quadrupole transition. The laser scheme for this part is depicted in (a).}
 \label{fig:Figure1}
 \end{figure}

  In this Letter, we propose a technique to collectively enhance the coupling between single
  photons and single ions using ion crystals. We consider a linear chain of ions inside an optical cavity (see
Fig. \ref{fig:Figure1}a). Strong coupling between a single ion and a
single photon is realized by collective enhancement of the coupling
of the photon with an ensemble of $N$ ions, given as $g_0\sqrt{N}$,
where $g_0$ is the vacuum Rabi frequency, i.e., the coupling between
a single ion and the incoming photon that enters the cavity. This means that the photon
state can be mapped onto a collective internal excitation of the
ions in the absorption process. Subsequently, this state will be
transferred to a phonon, i.e., a motional mode of the chain.
Finally, the phonon state will be mapped to a single-ion state.
Since the latter two couplings can in principle be accomplished with
arbitrarily strong laser fields, the collective $g_0\sqrt{N}$
coupling can dramatically improve the overall fidelity of
single-photon single-ion coupling.

We will show that using this mechanism it is possible to coherently
transfer with high fidelity an arbitrary internal state of a single
ion onto a single photon exiting the cavity, or an arbitrary state
of a single incoming photon onto an internal ionic state. This can be used 
for quantum coupling of single ion qubits in distant cavities, or
alternatively, for nonlinear quantum operations (quantum gates)
between single-photon qubits. In what follows we present an
analysis taking into account the inhomogeneous spacing of the
ions in a linear trap. This resulting inhomogeneous coupling between the cavity photon and
the ions has a sizable effect on the fidelities, and we
 suggest an approach for the
phonon-collective internal transition that will mimic the collective
internal-photon transition. The total fidelity thus can be maximized
by having the same relative coupling constants on each ion for both
transitions.

First, we briefly outline our approach for achieving strong single-photon single-ion coupling in an optical cavity with intermediate coupling to a single ion. The system consists of a string of
$N$ ions with a $\Lambda$-level internal structure as
shown in \fig{fig:Figure1}b, i.e., with ground states $\ket{0}$ and
$\ket{1}$ and an excited state $\ket{e}$. $g_0$ is the vacuum Rabi frequency
 on the $\ket{0}\rightarrow\ket{e}$ transition, and the transition $\ket{1}\rightarrow\ket{e}$ is driven by a classical field with Rabi
frequency $\Omega_1$ and homogeneous coupling to all ions.  We also assume a
quadrupole electric transition between the $\ket{0}$ and $\ket{1}$
states, with Rabi frequency $\Omega$.  The collective internal ion
excitation consists then of the Dicke-like states $\ket{\underline{\mbf
0}}=\ket{0_10_2\ldots 0_N}$ and $\ket{\underline{\mbf
1}}\propto\sum_ig_i\ket{0_10_2\ldots 1_i\ldots 0_N}$, etc., which takes into account the inhomogeneous coupling of the ions to the cavity field with coupling coefficients $g_i=g_0\sin(kz_i^0)$. Here $g_i$ is the
coupling of the cavity photon to ion $i$, where $z_i^0$ is the
equilibrium position of the ion, and $k$ the photon wave vector. The single-ion excitation is a
metastable state $\ket{1'}_i$, where $\ket{1'}$ can be the same or
different from $\ket{1}$. In our approach, in step I we map the
probe photon onto the collective ion excitation, $\ket{\underline{\mbf
0}}\rightarrow\ket{\underline{\mbf 1}}$, via a Stimulated Raman Adiabatic Passage
(STIRAP) process~\cite{STIRAP}. The probe transition is thus
collectively enhanced by $g_0\sqrt{N}$ \cite{Vuletic1}. Step II
consists of coupling this state to a phonon $\ket{\underline{\mbf
1};n_{pn}=0}\rightarrow\ket{\underline{\mbf 0};n_{pn}=1}$ (\fig{fig:Figure1}c), where $n_{pn}$ denotes the number of phonons in a selected mode, e.g.,
axial center-of-mass (COM),
via an adiabatic passage process, by using the quadrupole transition
coupling $\ket{0}$ and $\ket{1}$. In step III, the phonon is mapped
onto a single ion $j$, i.e. $\ket{\underline{\mbf
0};n_{pn}=1}\rightarrow\ket{0_1\ldots 1_j\ldots 0_N;n_{pn}=0}$.
 The main limitation to this scheme is the compromise between large $N$ for enhancing the $g_0\sqrt{N}$ coupling strength of the first step, and not too large $N$ such that the coupling to the phonons, scaling inversely with the mass of the system, is not reduced. We find about 40 ions to be both experimentally feasible, and providing sufficient coupling on both transitions.
 Thus single-photon single-ion coupling is achieved.

 The details of the scheme are most easily understood by describing the time reversed process, so we start by analyzing how to map a single-ion state onto a phonon state (step
   III).

In order to map the state of the $i$th ion, we propose to
use a Raman transition tuned to the red sideband, see
\fig{fig:Figure1}b,c, from $\ket{0_10_2\ldots1_i\ldots0_N;n_{\rm
pn}=0}$ to $\ket{\underline{\mbf 0};n_{\rm pn}=1}$. This process can be achieved
with fidelity $F_0=|\langle \psi_f|\psi(t)\rangle|^2$ larger than
0.99, where $|\psi_f\rangle$ and $|\psi(t)\rangle$ are the ideal
final state and the real state of the system at time $t$ respectively, when considering
the full Hamiltonian without rotating-wave approximation, and the
effect of the other passive phonon modes in the trap (see
below). The ion-to-phonon transfer is standard in trapped ion technology~\cite{LeibfriedEtAl}, and can be done with large fidelity:  The carrier Stark shift can be  canceled by detuning the laser, and the effect of the other modes is negligible for this Rabi frequency. Individual ion addressing is easy in this regime.

\begin{figure}[h] 
\includegraphics[width=0.8\linewidth]{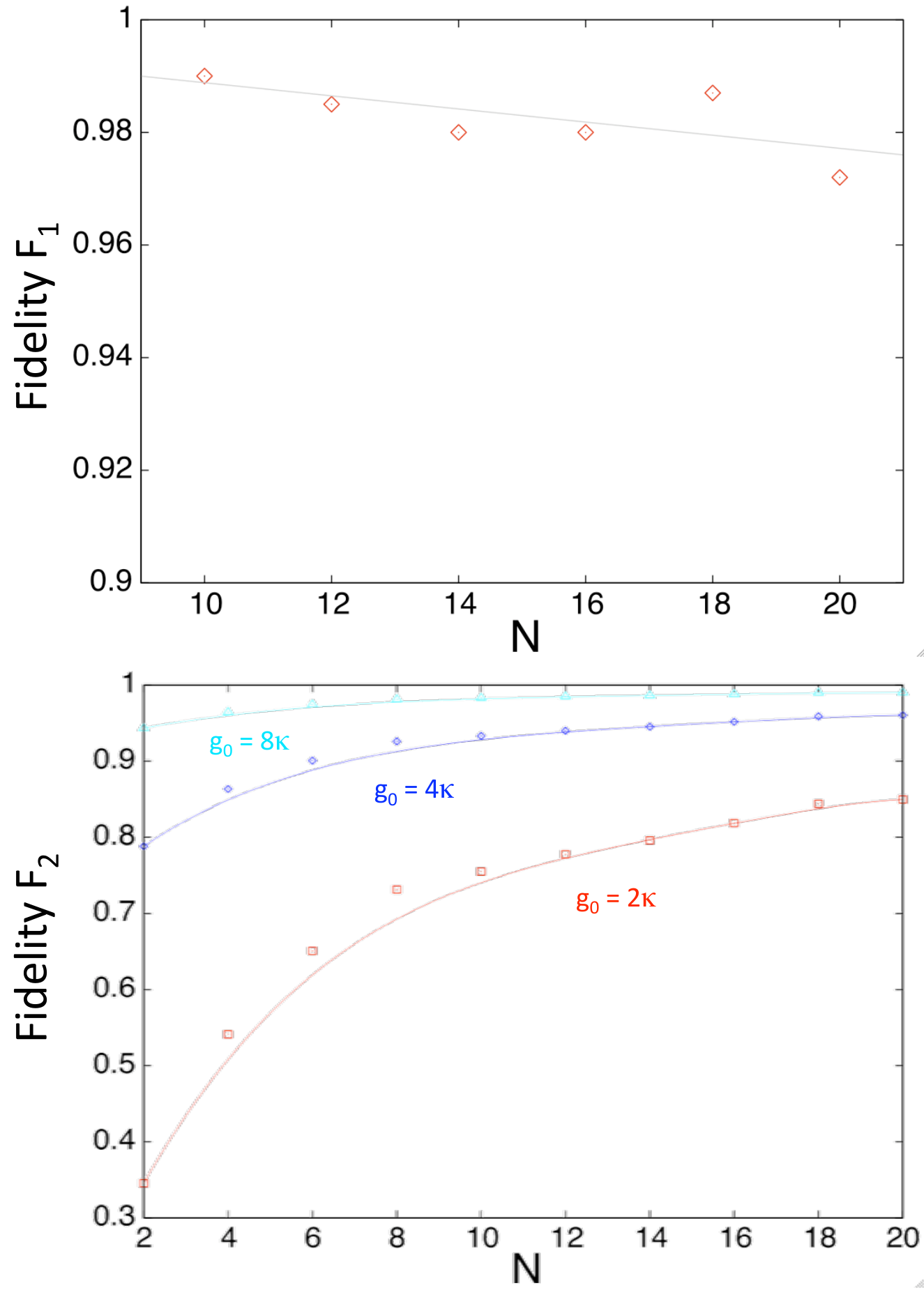}\hfill \\  
\caption{(a) Process fidelity $F_1$ for the transfer of one phonon onto one
collective excitation versus the number of ions, $N$. The
parameters used are: $\Omega_{\rm max}= 10^{-2}\omega$, $\eta=0.1$,
 $\tilde\eta=0.4$. The line is an average to guide the eye. (b) Fidelity $F_2$ for the
transfer of one collective excitation to one photon, versus the number of ions $N$.
The parameters used are $\Omega_1=50\kappa$,
$\Gamma=10\kappa$, $\Delta=0$. The lines are averages to guide the eye. 
}
                \label{fig:Figure2}
\end{figure}

For step II, in order to couple the phonon to a collective internal
ion excitation $\ket{\underline{\mbf 0};n_{pn}=1}\rightarrow\ket{\underline{\mbf
1};n_{pn}=0}$, we assume that states $\ket{1}$ and $\ket{0}$ are
connected by a metastable quadrupole electric transition. The reason
for considering this kind of transition here is that, as opposed to step III above, there is a phase matching condition: the
collective excitation has to fit the standing wave pattern of the
final photon in the cavity for optimal coupling, due to the ionic
inhomogeneous spacing. On the first glance, a simple Raman
transition would be the easiest choice here. Due, however, to the need to align the coupling laser(s) optimally with the modes, we
find a quadrupole transition made from two pairs of lasers works
best (depicted in \fig{fig:Figure1}a). Thus we consider a slightly non-coaxial standing laser field
configuration, with two pairs of lasers pairwise opposite in the $x$
direction, with one pair pointing rightwards, and the other one
pointing leftwards in the $z$ direction, see \fig{fig:Figure1}a. The
four lasers will give a joint field of
\begin{eqnarray}
&&A(x,z,t)
\propto \cos(k_x x)\sin(k z)\cos(\omega_0 t),
\end{eqnarray}
i.e., for all the ions with $x=0$ this gives $\sin(k z)\cos(\omega_0
t)$ with a correction in $x$ fluctuations that is quadratic in the
Lamb-Dicke parameter, and thus can be neglected. These lasers will
have frequency $\omega_0$, resonant with the quadrupolar transition.
The quadrupolar Hamiltonian contains the gradient of the field, such
that the resulting Hamiltonian will have a $\cos(k z)$ dependence
and will exhibit the typical standing wave sinusoidal pattern
needed, matching the cavity mode: $\cos (kz_i) \simeq \cos
(kz_{i}^0) - k\,\delta z_i \sin (kz_{i}^0)$, where $\delta z_i$ is
the fluctuation of the position operator $z_i$ around the
equilibrium position, $z_i^0$. This way, the ions will couple to the
phonon with equal relative couplings as to the photon, thus
maximizing the fidelity. Also, due to the angle $\theta$ the lasers
make with the cavity axis, their frequency will correspond to the
quadrupole electric transition, which usually is different from the
cavity frequency. The resulting complete Hamiltonian reads
\begin{eqnarray}
\label{Hphonon}
\lefteqn{H_1\,=\,
\omega b_{pn}^\dag b_{pn}+\sqrt{3}\omega \tilde b_{pn}^\dag \tilde b_{pn}+}\\
&&\Omega\sum_i(\sigma^+_i+\sigma^-_i)\Big\{\cos(kz_i^0)-
\sin(kz_i^0)\big[\frac{\eta}{\sqrt{N}}(b_{pn}+b_{pn}^\dag)\nonumber\\
&&+\frac{\tilde{\eta}}{\sqrt{N}}(\tilde b_{pn}+\tilde b_{pn}^\dag)\big]\Big\}+\Delta\sum_i\ket{1}_i\bra{1},\nonumber
\end{eqnarray}
where $b_{pn}^\dag$ is the main phonon mode, e.g., the COM mode,
that we will use for the protocol, and $\eta$ the single-ion
Lamb-Dicke parameter. In addition, the effect of the other modes
will be summed up in $\tilde b_{pn}^\dag$, whose frequency we take
as $\sqrt{3}\omega$ (the stretch mode frequency, nearest to the COM
one), by considering a larger Lamb-Dicke parameter for this mode.
Our estimates indicate that choosing $\tilde \eta=0.4=4\eta$ is
conservative enough to slightly overestimate the spurious effect of
the other modes. Accordingly, this is the value we will consider.
The other quantities are $\Omega_{\rm max}=0.01\omega$, where
$\Omega(t)$ has a Gaussian profile, and $\eta=0.1$. This operation
is optimal if one performs an adiabatic sweep of the detuning
$\Delta$ over the resonance with respect to the red sideband of the
$\ket{0}\rightarrow\ket{1}$ transition as plotted in
\fig{fig:Figure1}c. We take a maximum value of the detuning of $|\Delta-\omega|=8 \times10^{-3} \omega $,
and we consider a chirp with linear dependence on time. 
The process fidelity is given by
\[
F_1=\int d\alpha d\beta
\langle \psi_f|{\rm
Tr}[U|\psi_i\rangle\langle\psi_i|U^\dag]|\psi_f\rangle\delta(1-|\alpha|^2-|\beta|^2),
\]
where $|\psi_i\rangle=\alpha\ket{\underline{\mbf 0};n_{pn}=0}+\beta\ket{\underline{\mbf
0};n_{pn}=1}$ and $|\psi_f\rangle=\alpha\ket{\underline{\mbf
0};n_{pn}=0}+\beta\ket{\underline{\mbf 1};n_{pn}=0}$, $U$ is the evolution
operator associated with $H_1$, and the trace is taken over the
spurious mode. $F_1$ is depicted in \fig{fig:Figure2}a as a function
of the ion number $N$. While the main phonon mode, the COM mode, should be cooled down to
the ground state for performing the protocol, the remaining modes in
principle  may be cooled just to the Doppler limit. We verified that
there is no transfer of population from states $\ket{\underline{\mbf
0};n_{pn}=1}$ and $\ket{\underline{\mbf 1};n_{pn}=0}$ to other spurious phonon
modes, given that they are far off-resonant. In addition, the Stark
shifts induced by the spurious modes are negligible at the Doppler
limit. Accordingly, these modes need not be cooled down to the
ground state. For the matching to the cavity mode to work, the quadrupolar
interference pattern needs to be interferometrically stable with
respect to the cavity standing wave.

Finally, step I consists of coupling the collective ion excitation
to the cavity mode $a^\dag$, that subsequently will exit the cavity
\cite{CZKM}. This can be done using a straightforward Raman
transition in a STIRAP setup, between levels $\ket{\underline{\mbf 1};n_{ph}=0}$
and $\ket{\underline{\mbf 0};n_{ph}=1}$ (see \fig{fig:Figure1}b). The $\Omega_1$
laser is directed perpendicularly to the chain, such that no
additional phases are introduced. The Hamiltonian and master
equation for this system read

\begin{eqnarray}
\label{spontemissH}
H_2 & = & \sum_i\Big[ \Omega_1(\ket{e}_i\bra{ 1}+ \ket{1}_i\bra{e})+\\
&& g_0\sin(kz_i^0)(\ket{e}_i\bra{0} a
+H.c.)\Big]+\Delta\sum_i\ket{e}_i\bra{e},\nonumber\\
\dot{\rho} & = & -i[H_2,\rho]+\frac{\kappa}{2}(2a\rho a^\dag-a^\dag a\rho-\rho a^\dag a)+\label{spontemissrho}\\
&&\Gamma\sum_i(|1\rangle_i\langle e|\rho|e\rangle_i\langle 1|+\nonumber\\
&&|0\rangle_i\langle e|\rho|e\rangle_i\langle 0|-|e\rangle_i\langle e|\rho-\rho|e\rangle_i\langle e|),\nonumber
\end{eqnarray}
with  cavity decay rate $\kappa$. We can describe the transition using a STIRAP \cite{STIRAP} pulse. The fidelity $F_2$ of this process will be dominated by spontaneous emission from $\ket{e}$
\begin{eqnarray}
F_2&=&1-2\,\Gamma\int_0^\infty dt{\sum_i} \bra{e_i}\rho(t)\ket{e_i}.
\label{fidelityspontemiss}
\end{eqnarray}
%
We plot $F_2$ in \fig{fig:Figure2}b versus the number of ions, $N$.
Note that $F_2$ increases towards 1 with the number of
ions, $N$. This growth is especially significant for low $N$.  For $\omega/2\pi= 1 $MHz, the total transfer time of steps I-III is of about 2.3 ms for 20 ions. For the joint process fidelity of steps I-III, we obtain an optimal
value of $0.98$ for $N=18$ and $g_0=8\kappa$.

Having explained the basic setup, we can now discuss some potential extensions and applications. One such extension, two or more excitations, can be done in a straightforward manner. In this case, we consider a superposition state with $n$ photons $\alpha_0\ket{0}+\alpha_1\ket{1}+\ldots+\alpha_n\ket{n}$ which is mapped to the internal state $\alpha_0\ket{0_10_2\ldots 0_N}+\alpha_1\ket{0_1\ldots 1_{i_1}\ldots 0_N}+\ldots+\alpha_n\ket{0_1\ldots 1_{i_1}\ldots 1_{i_n}\ldots 0_N}$ where $i_1$ through $i_n$ denote $n$ definite (potentially neighboring) ions. The mapping process can then be achieved much the same way as described above, but the ``single'' ion--to--phonon step needs $n$ ions to be addressed individually by one or more lasers. In the phonon stage, $\ket{n}$ corresponds to $n_{pn}=n$, in the collective-ion-excitation stage to the Dicke state with $n$ excitations. The collective enhancement factor increases in this case to $\left(\begin{array}{c}N\\n\end{array}\right)^{\frac{1}{2}}$. 
For the coherent transfer from two excitations to two photons for
example, we obtain an $F_2$ (which is a lower bound to the
probability of no spontaneous emission in the two-excitation case)
from \eq{fidelityspontemiss} of 0.97 for 12 ions, with the
parameters $\Omega_1=50\kappa$, $\Delta=0$,
$\Gamma=10\kappa$, $g_0=8\kappa$.


An important practical consideration for the implementation of any
such operations concerns mapping of incident photon states into the
ion crystal. In the ideal limit, perfect mapping can be achieved,
via time reversal of the spin-photon mapping procedure discussed
above. In practice, the fidelity of this process will be limited by
finite optical depth, i.e., by the number of ions in the crystal.
The procedure of coupling photons into ensembles without and with
cavities have been investigated in detail previously
\cite{Review1,Review2,GardinerZoller}. In general, for optimally
chosen coupling strategies, the storage efficiencies will be similar
to the retrieval efficiencies \cite{Gorshkov2,Gorshkov3}.

There are multiple options for this scheme to be used as a photonic
gate, i.e., a nonlinearity that acts on two or more photons. In the
general case we will consider a superposition
$\ket{\psi_{ph}}=\alpha_0\ket{0}+\alpha_1\ket{1}+\alpha_2\ket{2}$ of
0, 1, and 2 photons for the incoming state, and the aim will be to
introduce a minus sign in the two-photon state, to get
$U^{(2)}\ket{\psi_{ph}}=\alpha_0\ket{0}+\alpha_1\ket{1}-\alpha_2\ket{2}$.
 A way to achieve that is based on transfer of the
state $\ket{\psi_{ph}}$ to a superposition of zero, one, and two
collective excitations, $\ket{\psi_{col}}=\alpha_0\ket{\underline{\mbf
0}}+\alpha_1\ket{\underline{\mbf 1}}+\alpha_2\ket{\underline{\mbf 2}}$, where $\ket{\underline{\mbf 2}}$
is the collective Dicke-like state with two excitations, $\ket{\underline{\mbf
2}}\propto\sum_{i\neq j}g_ig_j\ket{0_10_2\ldots 1_i\ldots 1_j\ldots
0_N}$.  Transferring subsequently to the
superposition of zero-, one-, and two-phonon states,
$\ket{\psi_{pn}}=\alpha_0\ket{n_{pn}=0}+\alpha_1\ket{n_{pn}=1}+\alpha_2\ket{n_{pn}=2}$,
and from it to the same superposition of single ion states, one
would get $\alpha_0\ket{0_10_2\ldots 0_N}+\alpha_1\ket{0_1\ldots
1_{i_1}\ldots 0_N}+\alpha_2\ket{0_1\ldots 1_{i_1} 1_{i_1+1}\ldots
0_N}$. We propose then to perform a standard two-qubit phase gate
upon the internal states of ions $i_1$ and $i_1+1$, introducing a
minus sign upon the $\alpha_2\ket{0_1\ldots 1_{i_1} 1_{i_1+1}\ldots
0_N}$ state. For example,
 the quantum-bus \cite{CiracZoller} or S{\o}rensen-M{\o}lmer \cite{Molmer} gates could be implemented in a straightforward manner.
 Subsequently, one would undo all the previous steps, and retrieve the
 photonic state with the minus sign incorporated upon the
 $\alpha_2\ket{n_{ph}=2}$ component of the state, thus completing
 the two-photon gate. A further possibility could be to consider the
 change in absorption of the cavity produced by the presence or
 absence of an excitation in the ions inside the cavity
 \cite{DuanKimble}. Thus, the phase of an incoming photon would be
 conditionally modified in the case where a former photon was
 previously absorbed.

From the above descriptions it is obvious that there is a large
number of applications possible with this scheme. This includes
quantum gates and quantum non-demolition measurements of
single-photon qubits \cite{KLM}, efficient entanglement purification
in ion-based quantum repeater schemes \cite{Repeater}, efficiency
enhancement in probabilistic ion entanglement schemes
\cite{Cabrillo}, as well as deterministic quantum gates between
distant trapped ions~\cite{CZKM}.

Finally, we point out that the here envisioned level scheme may be
attained in several presently possible ion trap setups such as
$^{40}$Ca$^+$ or $^{88}$Sr$^+$ trapped ions.

We thank M. C. Ba\~{n}uls, A. V. Gorshkov, O. Romero-Isart, T.
Sch\"{a}tz, and H. Schwager for fruitful discussions. We acknowledge funding from EU projects AQUTE and COMPAS, from DFG Forschergruppe 635, from the NSF CUA, DARPA QUEST, AFOSR MURI, and the Packard Foundation. S.F.Y.
acknowledges funding from NSF through award PHY-0970055.

\end{document}